\documentclass[fleqn,usenatbib,useAMS]{mnras}

\usepackage[T1]{fontenc}

\DeclareRobustCommand{\VAN}[3]{#2}
\let\VANthebibliography\thebibliography
\def\thebibliography{\DeclareRobustCommand{\VAN}[3]{##3}\VANthebibliography}

\usepackage{graphicx}	
\usepackage{amsmath}	
\usepackage{amssymb}	

\usepackage{newtxtext,newtxmath}

\title[eMSPs by resonant convection]{Eccentric millisecond pulsars by resonant convection}

\author[S. Ginzburg and E. Chiang]{
Sivan Ginzburg$^{1}$\thanks{E-mail: ginzburg@berkeley.edu}\thanks{51 Pegasi b Fellow.}
and Eugene Chiang$^{1,2}$
\\
$^{1}$Department of Astronomy, University of California, Berkeley, CA 94720-3411, USA\\
$^{2}$Department of Earth and Planetary Science, University of California, Berkeley, CA 94720-4767, USA
}

\date{Accepted XXX. Received YYY; in original form ZZZ}

\pubyear{2021}

\begin{document}
\label{firstpage}
\pagerange{\pageref{firstpage}--\pageref{lastpage}}
\maketitle

\begin{abstract}
Eccentric millisecond pulsars (eMSPs) with white dwarf companions exhibit orbital eccentricities orders of magnitude larger than predicted by turbulent convection in the white dwarfs' red giant progenitors. The orbital periods of eMSPs cluster around $P=$ 20 -- 30 d, remarkably close to the red giant convective eddy turnover time $t_{\rm eddy}$. We propose that the anomalously large eccentricities are resonantly driven by convective flows somehow made coherent when the turnover time matches the tidally locked red giant's spin period, which is also the tidal forcing period. Numerical simulations of rotating red giants and magnetic field studies of stars show some evidence for especially ordered flow patterns when the convective Rossby number $P/t_{\rm eddy}$ is of order unity. We show that resonant convection boosts eccentricities by a factor of $(t_{\rm nuc}/P)^{1/2}\approx 3\times 10^3$ over the random-walk values that characterize conventional MSPs, in good agreement with observations ($t_{\rm nuc}$ is the giant's nuclear burning time-scale). We also show how variations in the eddy turnover time arising from red giant metallicity variations can reproduce the observed effective width of the resonance, $\Delta P/P\approx 0.4$.
\end{abstract}




\section{Introduction}

Millisecond pulsars with white dwarf companions are thought to form when a red giant overflows its Roche lobe and transfers mass 
and spin angular momentum
to a neutron star. 
Nuclear shell burning around the red giant's inert helium core prescribes a relation between the core's mass and the giant's radius. Combining this relation with the Roche-lobe filling condition correlates the orbital period and the mass of the
helium
white dwarf that the giant leaves behind 
\citep{Joss1987,Savonije1987,PhinneyKulkarni94,Rappaport1995,TaurisSavonije99}.

\begin{figure}
\includegraphics[width=\columnwidth]{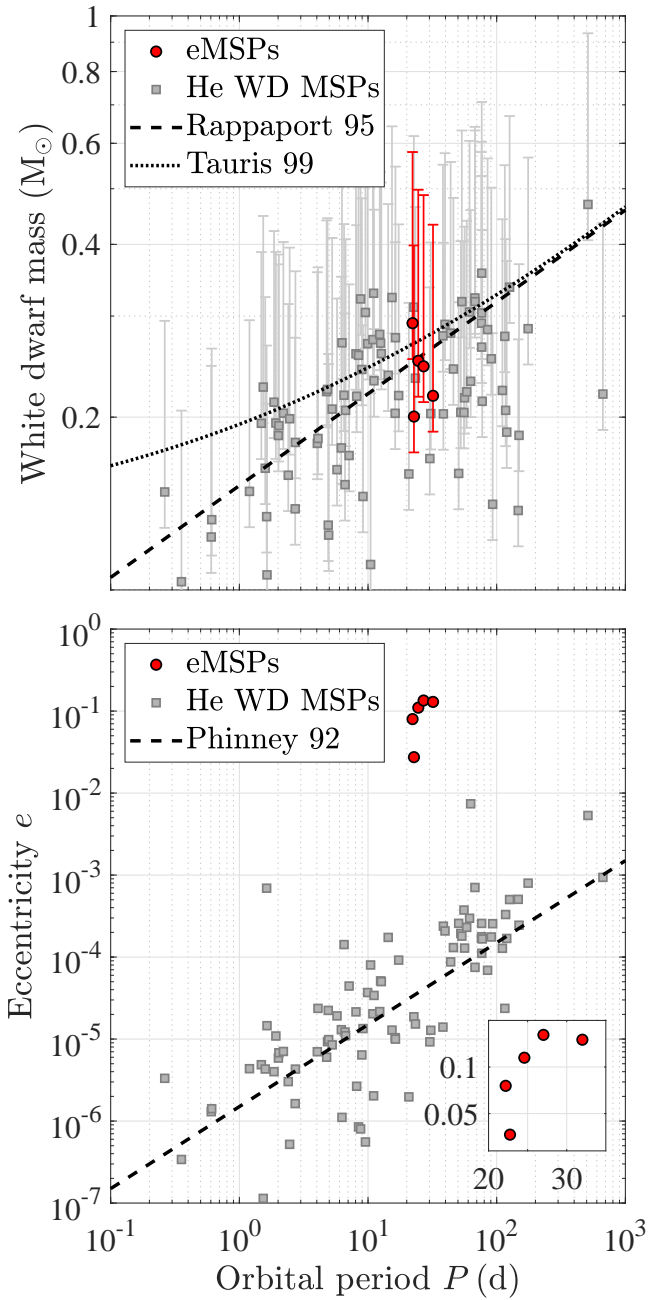}
\caption{Observed field millisecond pulsars \citep[spin periods shorter than 30 ms, e.g.][]{Lorimer2008} with helium white dwarf companions from the ATNF Pulsar Catalogue (\url{http://www.atnf.csiro.au/research/pulsar/psrcat}; \citealt{Manchester2005}). Eccentric systems ($e>10^{-2}$, eMSPs) are marked by red circles while other systems are marked by grey squares.
{\it Top panel:} Both eMSPs and conventional MSPs respect one mass--period relation (equation \ref{eq:period_mass}, with the normalization given by \citealt{Rappaport1995}), suggesting their common origin (namely Roche-lobe overflow of a red giant undergoing shell nuclear burning); \citealt{TaurisSavonije99} provide a more accurate computation at low masses. Points indicate the median white dwarf mass (inclination angle $i=60^\circ$) with error bars indicating the minimum ($i=90^\circ$) and 90 per cent probability ($i=25.8^\circ$) values. {\it Bottom panel:} The eMSPs deviate strongly from the general $e$--$P$ relation (equation \ref{eq:ecc_p}, with the normalization given by \citealt{Phinney1992})
within a narrow range of orbital periods $P \approx$ 20 -- 30 d (see the inset for a linear-scale zoom on the eMSPs).}
\label{fig:obs}
\end{figure}

The strong tides that the neutron star raises on the convective giant damp the orbital eccentricity down to a small but measurable value maintained by gravitational forcing from convective eddies. The giant's fluctuating quadrupole field stochastically perturbs the orbit until energy equipartition is reached between the epicyclic motion and the convective eddies \citep{Phinney1992}. The residual eccentricities typically range from $10^{-7}$
to $10^{-3}$, depending on the orbital period. 
In recent years, however, several field millisecond pulsars with much larger eccentricities have been discovered \citep{Champion2008,Bailes2010,Barr2013,Deneva2013, Camilo2015,Knispel2015,Octau2018}. These eccentric millisecond pulsars (eMSPs) have similar eccentricities of 0.03 -- 0.1 and similar orbital periods of 20 -- 30 d. Their white dwarf companions fall along the same mass--orbital period relation as describes conventional MSPs, suggesting that despite the dramatic differences in eccentricity, the same basic Roche-lobe overflow scenario applies to both populations \citep[see][and references therein]{Hui2018,HanLi2021,Jiang2021}.\footnote{We exclude PSR J1903+0327, which has a solar-mass main-sequence companion, and is therefore a clear exception \citep{Champion2008}. As this letter was nearing submission, \citet{Lorimer2021} announced the discovery of the mildly eccentric PSR J1146$-$6610 with an orbital period $P=62.8$ d and an eccentricity $e=7.4\times 10^{-3}$, which is just below our $10^{-2}$ eMSP threshold (we added the data from \citealt{Lorimer2021} to Fig. \ref{fig:obs}).}

We plot the observed systems in Fig. \ref{fig:obs}. We exclude pulsars in globular clusters, where dynamical interactions may have excited the eccentricity \citep{RasioHeggie95}. We also exclude systems with CO or ONeMg white dwarfs; these had more massive progenitors which may have undergone a phase of unstable mass transfer resulting in a qualitatively different evolution \citep{PhinneyKulkarni94}.    

Several ideas have been put forth to explain eMSPs. 
\citet{FreireTauris2014} argued that accretion-induced collapse of a massive (super-Chandrasekhar) white dwarf in a close binary could form a millisecond pulsar. In this scenario, the sudden release of binding energy induces a large orbital eccentricity. To avoid tidal circularization, the collapse must be delayed, presumably by rapid rotation, until 
after the companion's extended envelope has dissipated.  
\citet{Jiang2015} proposed a similar mechanism involving a phase transition from a neutron star to a strange quark star \citep[see also][]{Jiang2021}. Some eMSP masses, however, seem to be in tension with both of these scenarios, which predict rather specific pulsar masses \citep{Antoniadis2016,Barr2017,Stovall2019,Zhu2019}. \citet{Antoniadis2014} suggested that thermonuclear flashes due to unstable burning 
of residual gas
on top of the proto-white dwarf drive mass loss and lead to the formation of a circumbinary disc which can excite the system's eccentricity. The coupling between the disc and the binary in this scenario is sensitive to the disc's viscosity and to the possibility of its re-accretion \citep{Rafikov2016}. \citet{HanLi2021} argued that these  thermonuclear flashes would impulsively `kick'
the orbit by 
asymmetric mass loss, exciting eccentricity irrespective of the presence of a disc. Due to the observed narrow period range of eMSPs and the steep dependence of the orbital period on the white dwarf's mass (Fig. \ref{fig:obs}), any mechanism that relies on thermonuclear flashes must operate in an extremely fine-tuned mass range. \citet{HanLi2021} arbitrarily chose white dwarf masses of $0.268 - 0.281\,{\rm M}_\odot$ to fit the observations. This range is much narrower than found by numerical stellar evolution calculations; for example, \citet{Althaus2013} found $0.18 - 0.36\, {\rm M}_\odot$ 
white dwarfs to be susceptible to flashes; see also \citet{Istrate2016}. 
Even assuming an ad-hoc narrow range
of masses,
the scatter in the mass--period relation (top panel of Fig. \ref{fig:obs}) may produce eMSPs with a wider range of periods than observed.
It is easier to turn the argument around;
the fact that high eccentricities are observed only for a very narrow period range indicates that mass ejection during thermonuclear flashes is much weaker, or more symmetric, than assumed by \citet{HanLi2021}. 

In this letter, we present a new line of thinking to explain eMSPs that starts with their observed narrow period span. \citet{Phinney1992} found that the dominant convective eddies in the red giant's envelope turn over on a time-scale of $\approx 25$ d. This turnover time matches remarkably well the orbital periods of eMSPs. We therefore propose that a resonance between the convective turnover time and the orbital period is responsible for exciting the large eccentricities. This proposal rests on
the speculation that in resonance (and only in resonance), convective motions in the star cease being random and instead develop long-lived phase-coherent forcing fields.

The rest of this letter is organized as follows. In Section \ref{sec:evolution} we review the final stages in the red giant's evolution which determine the mass, period, and eccentricity of its white dwarf progeny. In Section \ref{sec:harmonic} we show how resonant convection can explain the eccentricities and period range of observed eMSPs. We summarize and give an outlook in Section \ref{sec:summary}.

\section{Red Giant Evolution}\label{sec:evolution}

In this section we follow the evolution of a red giant orbiting a neutron star up to the formation of a millisecond pulsar--white dwarf binary. A detailed discussion may be found in \citet{Phinney1992} and \citet{Rappaport1995}. Here we give a simplified and approximate re-derivation of the main scaling relations to provide context for our main result in Section \ref{sec:harmonic}.

\subsection{Radius}\label{sec:rad}

A star near the end of its life
fuses 
hydrogen in a shell surrounding an inert degenerate helium core 
of increasing 
mass $m_{\rm core}$ --- the ashes of nuclear burning. At small radii $r$ where the core's gravity dominates, the temperature $T$ is given by the virial theorem $kT\sim Gm_{\rm core}\mu/r$, where $G$ is the gravitational constant, $k$ is Boltzmann's constant, and $\mu$ is the mean molecular mass \citep[radiation pressure can be neglected for $m_{\rm core}\lesssim 0.5\,{\rm M}_\odot$, see][]{Kippenhahn2012}. Assuming a volumetric energy generation rate $\propto \rho^2 T^\nu$ from fusion, the luminosity is given by
\begin{equation}\label{eq:lum_gen}
    L\propto\int_{r_{\rm core}} \rho^2 T^\nu r^2{\rm d}r\propto r_{\rm core}^3\rho^2T^\nu.
\end{equation}
Since $T\propto r^{-1}$ and $\nu\gg 1$, the luminosity is dominated by a thin shell at the core radius $r_{\rm core}$ where the density $\rho$ and temperature $T$ are to be evaluated.
This luminosity is radiated outwards from the shell by diffusion
\begin{equation}\label{eq:lum_rad}
    L\propto\frac{r_{\rm core}^2T^4}{\tau}\propto\frac{r_{\rm core}^2T^4}{\kappa \rho r_{\rm core}}\propto \frac{r_{\rm core}T^4}{\rho},
\end{equation}
where $\tau$ and $\kappa$ denote the optical depth and opacity, respectively (the virial temperature dictates a density scale height $\sim r_{\rm core}$). We have assumed in the last proportionality a constant electron-scattering opacity appropriate for the high temperatures in the shell. By comparing equations \eqref{eq:lum_gen} and \eqref{eq:lum_rad} we find
\begin{equation}\label{eq:mass_lum}
    L\propto r_{\rm core}^{5/3}T^{(\nu+8)/3}\propto m_{\rm core}^{(\nu+8)/3}r_{\rm core}^{-1-\nu/3}\propto m_{\rm core}^{3+4\nu/9}\approx m_{\rm core}^9,
\end{equation}
where we used $T\propto m_{\rm core}/r_{\rm core}$, the mass--radius relation $r_{\rm core}\propto m_{\rm core}^{-1/3}$ for the degenerate (and significantly sub-Chandrasekhar) core, and $\nu\approx 13$ corresponding to the CNO cycle for the relevant temperatures. See \citet{RefsdalWeigert1970} and \citet{Kippenhahn2012} for a more rigorous derivation of equation \eqref{eq:mass_lum}.

The radius of the red giant, i.e. the radius of the hydrogen envelope surrounding the core, is given by
\begin{equation}\label{eq:mass_rad}
    r_{\rm env}\propto \left(\frac{L}{T_{\rm eff}^4}\right)^{1/2}\propto m_{\rm core}^{9/2},
\end{equation}
where $T_{\rm eff}$ is the giant's effective temperature, which remains approximately constant (because of the strong dependence of the photospheric ${\rm H}^-$ opacity on $T_{\rm eff}$) along the Hayashi track. See \citet{Rappaport1995} for a more accurate version of equation \eqref{eq:mass_rad}.

Equation \eqref{eq:mass_rad} shows that as nuclear burning adds mass to the core, the  envelope expands. Eventually the giant overflows its Roche lobe. If the ensuing mass transfer is stable, the orbit adjusts such that the giant keeps marginally filling its Roche lobe as it expands further. The hydrogen envelope is gradually depleted by both fusion and Roche-lobe overflow. When its mass $m_{\rm env}$ becomes comparable to the mass of the burning shell --- typically around $10^{-2}m_{\rm core}$ --- equation \eqref{eq:mass_rad} 
breaks down (the shell and envelope can no longer be considered separately) and the envelope begins contracting \citep{RefsdalWeigert1969,RefsdalWeigert1970,RefsdalWeigert1971,Phinney1992}. At this point the star detaches from its Roche lobe and its orbital period is set at 
\begin{equation}\label{eq:period_mass}
    P\propto\left(m_{\rm core}^{-1/3}r_{\rm env}\right)^{3/2}\propto m_{\rm core}^{25/4}\approx m_{\rm core}^6,
\end{equation}
where we used equation \eqref{eq:mass_rad} and assumed that the neutron star is much heavier than the core; see \citet{Rappaport1995} and \citet{TaurisSavonije99} for a more accurate calculation. The envelope keeps fuelling the shell and contracting, until the star completes its evolution into a white dwarf with mass $m_{\rm core}$. Thus, equation \eqref{eq:period_mass} for the orbital period at the moment of Roche-lobe detachment sets the final mass--period relation (Fig. \ref{fig:obs}).

\subsection{Eccentricity}\label{sec:ecc}

The circularization of MSP--proto white dwarf orbits is calculated by generalizing the \citet{Zahn1977} tidal theory to the core--envelope structure of red giants \citep[see][]{Phinney1992}. We assume for simplicity $m_{\rm env}\ll m_{\rm core}\ll M$, where $M$ is the mass of the neutron star. 
The eccentricity tide raised on the giant by the neutron star oscillates with 
orbital period $P$ and
amplitude $\xi\sim r_{\rm env}(M/m_{\rm core})(r_{\rm env}/a)^3e$, where $a$ and $e\ll 1$ are the orbital semi-major axis and eccentricity. The energy in these oscillations $E_{\rm tide}\sim m_{\rm env}(\xi/P)^2$ is dissipated by turbulent convective eddies on the eddy turnover time $t_{\rm eddy}\sim r_{\rm env}/v_{\rm eddy}$, where $v_{\rm eddy}$ is the convective velocity. We assume here that the envelope is fully convective
(ignoring the thin radiative shell above the helium core), such that the largest and dominant eddies are of scale $r_{\rm env}$ \citep{Phinney1992}. After an energy
$E_e=e^2GMm_{\rm core}/(2a)$ has been dissipated, the orbit is circular (assuming orbital angular momentum is conserved in the process). The circularization time-scale is therefore
\begin{equation}\label{eq:t_circ}
    t_{\rm circ}\sim t_{\rm eddy}\frac{E_e}{E_{\rm tide}}\sim t_{\rm eddy}\frac{m_{\rm core}}{m_{\rm env}}\left(\frac{m_{\rm core}}{M}\right)^2\left(\frac{a}{r_{\rm env}}\right)^8,
\end{equation}
where we used $P^2\sim a^3/(GM)$. 
The convective velocity adjusts to carry the nuclear energy flux: $\rho v_{\rm eddy}^3\sim L/r_{\rm env}^2$ for envelope density $\rho\sim m_{\rm env}/r_{\rm env}^3$. Then
\begin{equation}\label{eq:t_eddy}
    t_{\rm eddy}\sim\frac{r_{\rm env}}{v_{\rm eddy}}\sim r_{\rm env}\left(\frac{Lr_{\rm env}}{m_{\rm env}}\right)^{-1/3}\sim\left(\frac{m_{\rm env}r_{\rm env}^2}{L}\right)^{1/3}.
\end{equation}

When the giant fills its Roche lobe, $r_{\rm env}$ is large and thus $t_{\rm circ}$ is short according to equation \eqref{eq:t_circ}; eccentricities are efficiently damped. Nevertheless, a non-zero residual eccentricity is maintained by energy equipartition between the epicyclic motion of the orbit
and the motions of convective eddies: $E_e\sim m_{\rm env} v_{\rm eddy}^2$ \citep{Phinney1992}. Hence
\begin{equation}\label{eq:ecc_p}
    e\propto\left(\frac{m_{\rm env}}{m_{\rm core}}\right)^{1/2}P^{1/3}v_{\rm eddy}\propto m_{\rm env}^{1/6}m_{\rm core}^6\propto P,
\end{equation}
where we substituted $v_{\rm eddy}$ from equation \eqref{eq:t_eddy} and used equations \eqref{eq:mass_lum}--\eqref{eq:period_mass}. We follow \citet{Phinney1992} and neglect the weak dependence on $m_{\rm env}$ (which itself depends on $m_{\rm core}$) in the last proportionality of equation \eqref{eq:ecc_p}. When the giant detaches from its Roche lobe and begins contracting, $t_{\rm circ}$ lengthens according to equation \eqref{eq:t_circ}. When $r_{\rm env}$ decreases to about half the size of the Roche lobe, $t_{\rm circ}$ becomes longer than the envelope's contraction time, which is set by nuclear burning (see Section \ref{sec:rad}). From this point on, the giant is unable to damp its eccentricity, which `freezes' at the value set by equation \eqref{eq:ecc_p}. Equation \eqref{eq:ecc_p} thus sets the final eccentricity of the white dwarf, explaining the observed period--eccentricity relation for conventional helium white dwarf--MSP binaries (Fig. \ref{fig:obs}). 

\section{Resonance}\label{sec:harmonic}

The residual eccentricity $e$ can be derived more directly. 
The radial excursion $x$ away from a circular orbit of radius $a$ can be modelled as a harmonic oscillator with frequency $\Omega=2\upi/P$ \citep{Phinney1992}:
\begin{equation}\label{eq:harmonic}
    \ddot{x}+\frac{\dot{x}}{t_{\rm circ}}+\Omega^2 x=f(t),
\end{equation}
where $f$ is the driving acceleration 
felt by the red giant (more precisely, by the reduced mass).
For $f=0$ and $t_{\rm circ}\to\infty$, equation \eqref{eq:harmonic} describes
a Keplerian orbit with a constant radial amplitude $x = ae$ (to first order in $e$).

In our case, the driving $f$ originates from the fluctuating gravitational potential of the giant's convective envelope. In convection, heated fluid elements rise buoyantly due to their lower density, with a velocity given by $v_{\rm eddy}^2\sim (Gm_{\rm core}/r_{\rm env}^2)(\delta \rho/\rho)r_{\rm env}$, where $\delta \rho$ is the density fluctuation.\footnote{\citet{Phinney1992} generalizes the calculation to scale heights $H<r_{\rm env}$ while we focus only on the dominant eddies with $H\sim r_{\rm env}$. The two calculations differ by a dimensionless factor, and we use the \citet{Phinney1992} normalization in Fig. \ref{fig:obs}.} Conservation of mass and momentum fix the giant's gravitational monopole and dipole moments, such that the lowest fluctuating term is the quadrupole.   
The acceleration that the
neutron star
experiences due to the giant's density fluctuations is therefore
$(Gm_{\rm env}r_{\rm env}^2/a^4)(\delta\rho/\rho)$. The giant's 
back-reactive acceleration 
is larger by a factor of $M/m_{\rm core}$ (we again approximate for simplicity $m_{\rm env}\ll m_{\rm core}\ll M$):
\begin{equation}\label{eq:f}
    f\sim\frac{Gm_{\rm env}r_{\rm env}^2}{a^4}\frac{M}{m_{\rm core}}\frac{\delta\rho}{\rho}\sim \frac{m_{\rm env}}{m_{\rm core}}\frac{M}{m_{\rm core}}\left(\frac{r_{\rm env}}{a}\right)^4\frac{v_{\rm eddy}^2}{r_{\rm env}}.
\end{equation}

\citet{Phinney1992} models the convective eddies as white noise at low frequencies (i.e. long time-scales $t\gg t_{\rm eddy}$),
with $f$ as a stochastic function.
In one eddy turnover time, 
the eddies change the giant's epicyclic velocity $v_e\equiv\Omega x=\Omega ae$ by $\sim$$ft_{\rm eddy}$, and over many turnover times $v_e$ grows by random walking:
\begin{equation}\label{eq:random_walk}
    v_e\sim f t_{\rm eddy}\left(\frac{t_{\rm circ}}{t_{\rm eddy}}\right)^{1/2}=f\left(t_{\rm eddy}t_{\rm circ}\right)^{1/2},
\end{equation}
with $t_{\rm circ}$ limiting the time over which the eccentricity may grow. Using equations \eqref{eq:t_circ}, \eqref{eq:f}, and \eqref{eq:random_walk}, we find that the orbital epicyclic energy
\begin{equation}\label{eq:equipar}
    E_e=\frac{1}{2}m_{\rm core}v_e^2\sim m_{\rm core}f^2 t_{\rm eddy}t_{\rm circ}\sim m_{\rm env} v_{\rm eddy}^2 
\end{equation}
is in equipartition with the eddy kinetic energy, as assumed in the derivation of the $e$--$P$ relation in equation \eqref{eq:ecc_p} (\citealt{Phinney1992} points out that this is a manifestation of the fluctuation--dissipation theorem).

The eccentricity implied by equations \eqref{eq:random_walk} and \eqref{eq:equipar} explains well the observations (see Fig. \ref{fig:obs} and Section \ref{sec:ecc}) except for the anomalous eMSPs, which cluster at orbital periods of $\approx$ 20 -- 30 d. This narrow range coincides with resonance: here the orbital period equals the dominant eddy's turnover time,  $P \sim t_{\rm eddy}$. The turnover time can be estimated to order-of-magnitude using equation \eqref{eq:t_eddy} for $m_{\rm env}\approx 7\times 10^{-3}\,{\rm M}_\odot$ and $T_{\rm eff}\approx 7\times 10^3\,{\rm K}$ \citep[][for $m_{\rm core}\approx 0.26\,{\rm M}_\odot$]{RefsdalWeigert1969} :
\begin{equation}\label{eq:teddy_teff}
    t_{\rm eddy}\sim \left(\frac{m_{\rm env}}{\sigma T_{\rm eff}^4}\right)^{1/3}
\end{equation}
where $\sigma$ is the Stefan-Boltzmann constant. Note that $T_{\rm eff}$ is somewhat above the Hayashi line because it is evaluated when the envelope has contracted to half its maximum value (see Section \ref{sec:ecc}). In a more careful calculation that kept track of order-unity coefficients, \citet{Phinney1992} found that the resonance period is $P=t_{\rm eddy}\approx 25\,{\rm d}$ --- exactly where the eMSPs are being found more than two decades later. 

To good approximation, there are just one or a few dominant eddies, as the scale height in the convective envelope's interior is of order $r_{\rm env}$.
We now make the ansatz that at resonance these eddies do not randomly change direction and decorrelate every turnover time, as assumed by \citet{Phinney1992}. Rather, we assume they persist as long-lived vortices generating a quadrupole moment that oscillates coherently and not stochastically. 
This ansatz appears supported by the fact that, at resonance, the tidally locked \citep{Phinney1992} red giant's Rossby number ${\rm Ro}\equiv P/t_{\rm eddy}=1$, 
signalling the importance of rotation in determining convective flow patterns. 
Three-dimensional numerical simulations
suggest
that at ${\rm Ro} \sim 1$, convection organizes into coherent large-scale structures \citep[e.g.][]{BrunPalacios2009}; see Section \ref{sec:summary} for a more detailed discussion.
It might also be that coherent circulation patterns are resonantly enforced by the oscillating tidal field of the orbiting neutron star. Assuming $f$ is harmonic, equation \eqref{eq:harmonic} can be solved as a standard damped oscillator driven at frequency $\omega= 2\upi/t_{\rm eddy}$. The amplitude is given by 
\begin{equation}\label{eq:driven}
    x=\frac{f}{\sqrt{\left(\Omega^2-\omega^2\right)^2+\left(\omega/t_{\rm circ}\right)^2}}.
\end{equation}
At resonance $\omega=\Omega$, equation \eqref{eq:driven} predicts an epicyclic velocity
\begin{equation}\label{eq:ve_res}
    v_e^{\rm res}=\Omega a e=\Omega x=ft_{\rm circ},
\end{equation}
assuming the eddies coherently perturb the orbit over $t_{\rm circ}$ (the maximal time over which the eccentricity may grow).

Comparing equations \eqref{eq:random_walk} and \eqref{eq:ve_res}, we see that the eccentricity at resonance is enhanced by a factor of
\begin{equation}\label{eq:enhance}
    \frac{e^{\rm res}}{e}=\frac{v_e^{\rm res}}{v_e}\sim \left(\frac{t_{\rm circ}}{t_{\rm eddy}}\right)^{1/2}\sim\left(\frac{t_{\rm nuc}}{P}\right)^{1/2}\approx 3\times 10^3,
\end{equation}
where $t_{\rm eddy}=P\approx 25\,{\rm d}$ at resonance. Extending our discussion in Section \ref{sec:evolution}, $t_{\rm circ}$ at the moment the eccentricity `freezes out' equals the nuclear burning time $t_{\rm nuc}$ over which the envelope contracts. From \citet{RefsdalWeigert1969}, $t_{\rm nuc}\sim 10^6\,{\rm yr}$ at freeze-out, when the giant has contracted to about half the Roche lobe radius. \citet{Phinney1992} cites a somewhat longer $t_{\rm nuc}\sim 10^7\,{\rm yr}$ appropriate for a giant that still fills its Roche lobe. All these numerical estimates are made for $m_{\rm core}\approx 0.26\,{\rm M}_\odot$, which fits the observed eMSP masses as well as the theoretical mass--period relation for the resonance period (Fig. \ref{fig:obs}). The enhancement given by equation \eqref{eq:enhance} matches the deviation of the eMSPs from the \citet{Phinney1992} relation at the same orbital period ($e^{\rm res}\approx 10^{-1}$ instead of $e\approx 3\times 10^{-5}$). 

As seen in equation \eqref{eq:driven}, the width of the resonance is $\Delta\Omega\sim t_{\rm circ}^{-1}$, or equivalently
\begin{equation}
    \frac{\Delta P}{P}\sim \frac{P}{t_{\rm circ}}\sim\frac{P}{t_{\rm nuc}}\approx 10^{-7}.
\end{equation}
Staying within such a narrow resonance may not be a problem insofar as a suitably wide range of eddy forcing frequencies (centred on $t_{\rm eddy}^{-1}$) is exhibited by a given red giant at any time --- the variation between the specific turnover times of individual eddies effectively widens the resonance. As the red giant evolves during freeze-out and the mean $t_{\rm eddy}$ changes somewhat according to equation \eqref{eq:teddy_teff}, there should always be some power at frequency $P^{-1}$ to drive the orbit, even though the peak of the power spectrum may have shifted somewhat.
Can we reproduce the $\approx$ 40 per cent spread in observed eMSP orbital periods (Fig. \ref{fig:obs})? This range could arise from variations in $t_{\rm eddy}$ due to differences in the compositions of main sequence progenitors. 
On the Hayashi track, the strong dependence of  ${\rm H}^-$ opacity on metallicity results in a variation of about 30 per cent in $T_{\rm eff}$ for different compositions \citep[fig. 24.3 of][]{Kippenhahn2012}. Using equation \eqref{eq:teddy_teff}, this leads to a $\approx$ 40 per cent variation in $t_{\rm eddy}\propto T_{\rm eff}^{-4/3}$. These considerations also reproduce the factor of $\approx 2$ scatter in $P$ at fixed $m_{\rm core}$ as found by \citet[][see their fig.~4a]{TaurisSavonije99}; from equations \eqref{eq:mass_rad} and \eqref{eq:period_mass}, $P\propto r_{\rm env}^{3/2}\propto T_{\rm eff}^{-3}$. 
Note that metallicity-induced variations in
$t_{\rm eddy}$ broaden the range of potentially resonant $P$ but also leave some systems out of resonance, and indeed Fig. \ref{fig:obs} shows that at fixed $P$ between 20 -- 30 d there are both eMSPs and conventional MSPs. 

\section{Summary and Discussion}\label{sec:summary}

Helium white dwarf companions to millisecond pulsars provide a fossil record of the progenitor red giant phase, when the white dwarf 
was still growing within 
the giant's extended convective envelope. The mass--period relation exhibited by these white dwarf companions is set by nuclear shell burning atop the nascent degenerate core \citep[e.g.][]{Rappaport1995}, and their tiny but non-zero orbital eccentricities are a relic of gravitational forcing by convective eddies in the giant's envelope \citep{Phinney1992}.
The discovery of several millisecond pulsars with anomalously large eccentricities (eMSPs) challenges this picture, as their eccentricities, on the order of 0.1, should have been damped by dissipative 
tides raised in
the envelope. At the same time, eMSPs obey the same mass--period relation of the more circular systems, suggesting all share the same basic lineage. 

The narrowness of the range of orbital periods exhibited by eMSPs suggests the action of a resonance. We propose that the eccentricities of eMSPs, like their conventional MSP counterparts, were excited by convective eddies during the red giant phase, with the qualitative difference that the eddies in eMSP systems acted in resonance with the orbital motion. The orbital period $P$ equals the eddy turnover time $t_{\rm eddy}$ for $P\approx 25\,{\rm d}$ \citep{Phinney1992} --- squarely in the range of eMSP periods. 
This resonance
can enhance the eccentricity only if the eddies form 
long-lived structures that
circulate coherently. 
We hypothesize that 
the giant's quadrupole moment oscillates periodically when $P \sim t_{\rm eddy}$, but varies stochastically when far from resonance.

Since the red giant is tidally locked \citep{Phinney1992}, its spin period 
is equal to the orbital period $P$. Thus when $t_{\rm eddy} = P$ on resonance, the giant's 
convective Rossby number
${\rm Ro}\equiv P/t_{\rm eddy}\sim 1$. 
This value (up to order-unity adjustments) 
has been associated with the emergence of large scale coherent vortices \citep[e.g.][and references therein]{Guervilly2014,LinJackson2021}. In their three-dimensional simulations of a rotating red giant, \citet{BrunPalacios2009} find a transition from disordered eddies to coherent `dipolar' convection at ${\rm Ro}\sim 1$ (their fig. 5).
The number of convection cells at mid-depth is about ten at most (even in the disordered case), consistent with our
order-of-magnitude calculation. \citet{Garraffo2018} offer other evidence for coherent convection at order-unity ${\rm Ro}$. By analysing Zeeman--Doppler images and rotation periods for 
main-sequence stars, they conclude that the magnetic `complexity' (the number of significant terms in the multipole expansion) reaches a minimum at ${\rm Ro}\approx 0.1$ (their fig. 1), suggesting  
a corresponding minimum in the complexity of the underlying convective flow (note that there are order-unity differences between their definition of ${\rm Ro}$ and that of \citealt{BrunPalacios2009}). 
At this minimum, \citet{Garraffo2018} find a nearly dipolar field. 

Coherent circulation might also be maintained by resonant forcing from the tidal field of the neutron star.
The interplay of convection and tides is sensitive to $P/t_{\rm eddy}$ but the picture remains hazy.
Most numerical simulations to date are limited to small domains of the stellar envelope and neglect rotation \citep{Duguid2020}. The discovery of eMSPs motivates global numerical simulations of convection in tidally locked red giants on non-circular orbits with ${\rm Ro}\sim 1$ (see \citealt{Jones2011} for a discussion of the computational challenge).

Though unproven and barely tested, the hypothesis that at resonance eddies can persist coherently for long time-scales would seem to neatly explain
eMSPs. We have shown that resonant convection
enhances the
orbital eccentricity by a factor of $(t_{\rm circ}/t_{\rm eddy})^{1/2}=(t_{\rm nuc}/P)^{1/2}\approx 3\times 10^3$ compared to the equipartition value derived from the off-resonance random walk (the system's eccentricity is set when the tidal circularization time $t_{\rm circ}$ equals the nuclear burning time $t_{\rm nuc}$). Such an enhancement is just what is needed to reproduce measured eMSP eccentricities of $e\approx 10^{-1}$. 
We also 
argued 
that the observed width of the resonance  $\Delta P/P\approx 0.4$ could reflect 
variations 
in $t_{\rm eddy}$ stemming from metallicity variations in red giant progenitors. 
This predicts that shorter-period eMSPs 
originate 
from lower-metallicity giants  
and therefore 
have slightly above-average white dwarf masses 
for their period 
\citep{TaurisSavonije99}.

The recently discovered mildly eccentric ($e=7.4\times 10^{-3}$) PSR J1146$-$6610 \citep{Lorimer2021} has an orbital period $P=62.8$ d, which is about twice that of the eMSPs (Fig. \ref{fig:obs}). It is not clear whether this system implies a wider resonance in our scenario \citep[or a higher-order resonance, as suggested by][]{Lorimer2021}, or whether it represents the high end of the scatter in non-resonant systems. 
\section*{Acknowledgements}

We thank Matthew Browning, Yohai Kaspi, Xiang-Dong Li, Duncan Lorimer, Thomas Tauris, and the anonymous reviewer for helpful discussions and comments. SG is supported by the Heising-Simons Foundation through a 51 Pegasi b Fellowship.

\section*{Data Availability}

The data underlying this article were derived from the ATNF Pulsar Catalogue \citep{Manchester2005}, version 1.64 (November 2020): \url{http://www.atnf.csiro.au/research/pulsar/psrcat}. 



\bibliographystyle{mnras}
\input{eMSP.bbl}




\bsp	
\label{lastpage}
\end{document}